\documentclass{iaus}
\usepackage{graphicx}
 
\title[S262 ~~Stellar Populations - Planning for the Next Decade] 
{The Mg/Fe characterization of the \\ MILES library for stellar populations studies}
 
\author[A. Milone, A.E. Sansom \& P. S\'anchez-Bl\'azquez]   
{Andr\'e Milone$^{1,2}$,
Anne E. Sansom$^2$ \and Patricia S\'anchez-Bl\'azquez$^3$}
 
\affiliation{$^1$Divis\~ao de Astrof\'\i sica, Instituto Nacional de Pesquisas Espaciais, Brazil \\
email: {\tt acmilone@das.inpe.br} \\[\affilskip]
$^2$Jeremiah Horrocks Institute, University of Central Lancashire, United Kingdom \\
$^3$Instituto de Astrof\'\i sica de Canarias, Spain}
 
\pubyear{2009}
\volume{262}  
\pagerange{001--002}
\jname{Stellar Populations - Planning for the Next Decade}
\editors{G.R. Bruzual, \& S. Charlot, eds.}
\begin{document}
 
\maketitle
 
\begin{abstract}
We have obtained [Mg/Fe] for around 77\% of the stars of the MILES library of stellar spectra
in order to include this important information into simple stellar population (SSP) models.
The abundance ratios, which were carefully calibrated to a single uniform scale,
were obtained through a compilation from high spectral resolution works plus
robust spectroscopic analysis at medium resolution. The high resolution data provided
an extensive control sample. Average uncertainties (0.06 and 0.12 dex for the high and
medium resolution samples respectively) and the good coverage of the stars with [Mg/Fe]
over the MILES's parameter space will permit us to semi-empirically build up new SSP models
with accurate $\alpha$-enhancements for ages older than 1 Gyr.
This will open new prospects for evolutionary stellar population synthesis.
 
\end{abstract}
 
\firstsection 

\section{Introduction}
 
One current limitation of the SSP models that are based on empirical stellar specta
is that they rely on Galactic stars whose detailed chemical properties are not well known.
For example, the stars in the different components of
the Galaxy show distinct patterns of alpha-elements
reflecting different star formation histories.
The models usually only take into account the iron abundance, but
stellar spectra may change considerably if the ratios between other
elements and Fe are different.
 
Accounting for well known element abundances will make an
empirical spectral stellar library particularly useful for modeling
stellar populations with different formations.
In this work, we present Mg/Fe ratios for 758 MILES 
\cite[S\'anchez-Bl\'azquez {\it et al.} (2006)]{Sanchez-Blazquez06}'s stars.
The values are a combination of a literature compilation and 
our own measurements.

\section{Mining Mg abundances and the [Mg/Fe] scale}
In a first step, we obtained published Mg abundances 
from high resolution spectra of the MILES's stars.
We adopted the
\cite[Borkova \& Marsakov (2005)]{BorkovaMarsakov05}'s
compilation for defining a consistent scale of [Mg/Fe]
that was based on weighted averages.
Calibration of the ratios from a given work to this scale
relies on a statistically representative linear relation
by using stars in common between two samples.
 
We used 219 stars from that compilation and
also collected abundance ratios for another 89 stars from high resolution analyses,
thus covering $\sim$1/3 of the library having 248 dwarfs and 60 giants.
Sixteen stars from duplicated sources helped us to evaluate the calibration process
as well as estimate the final errors of [Mg/Fe], which are around 0.06 dex.
This step was extremely useful in defining a representative reference sample
for calibrating our measurements.
 
\section{Mg abundances measured at medium resolution}
 
Our Mg abundance measurements were based on a spectral synthesis
computed with the MOOG LTE code
(Sneden 2002, http://verdi.as.utexas.edu/)
applied to MILES spectra through an automatic process
using both pseudo equivalent widths and line profile fits
of two Mg features
(MgI$\lambda$5528.40 {\AA} and the Mg b triplet).
We adopted linearly interpolated model atmospheres
over the most recent MARCS grid
\cite[(Gustafsson {\it et al.} 2008)]{Gustafsson08}
and accurate atomic/molecular
line lists in order to compute a reliable set of synthetic spectra
for five values of [$\alpha$/Fe], for each star.
Our measured [Mg/Fe] values cover about 46\% of the MILES stars
(152 dwarfs and 298 giants),
with errors ranging from 0.08 to 0.15 dex (weighted average of 0.12 dex).
See them as a function of [Fe/H] in Figure 1.
 
\begin{figure}[t]
\begin{center}
\includegraphics[width=8.4cm, angle=-90]{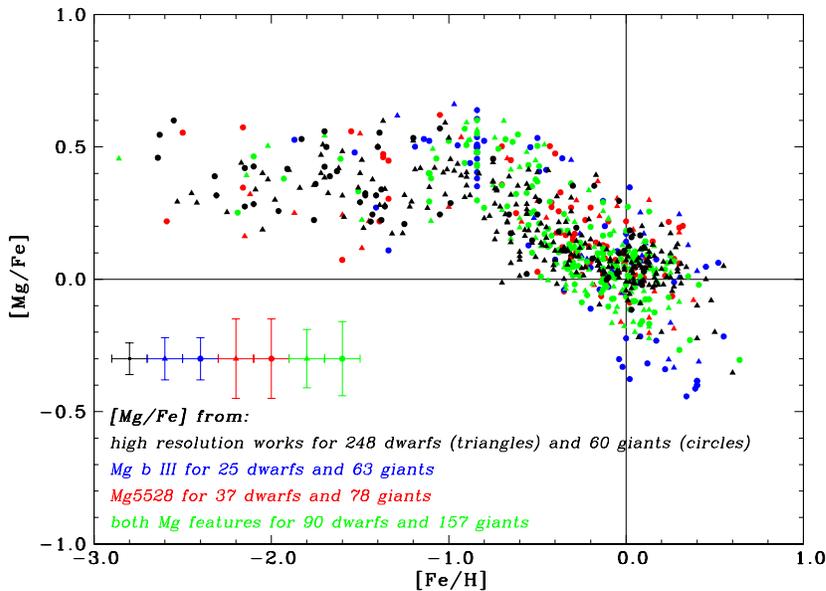}
\caption{[Mg/Fe] {\it vs.} [Fe/H]: high and medium resolution
measurements onto uniform scales.}
\label{fig1}
\end{center}
\end{figure}

\section{Results and further applications}
 
We have obtained [Mg/Fe] for $\sim$3/4 of MILES
(400 dwarfs and 358 giants, respectively 74\% and 82\% of them).
The results cover the MILES's parameter space quite well.
Specifically, we recovered at medium resolution the Mg abundances
for a lot of sub-giants and red giants
although a lack still remains on the stars cooler than 3500~K.

We will use these abundances to compute
semi-empirical SSPs (older than 1 Gyr) with different [Mg/Fe].
These refined SSP models will be tested against globular clusters.

\begin{acknowledgements}
 
A. Milone thanks the Brazilian foundations CAPES and FAPESP.
 
\end{acknowledgements}


\begin{thebibliography}{}
 
\bibitem[Borkova \& Marsakov (2005)]{BorkovaMarsakov05}
{Borkova, T.V. \& Marsakov, V.A.} 2005,
\textit{AZh}, 82, 453
 
\bibitem[Gustafsson {\it et al.} (2008)]{Gustafsson08}
{Gustafsson {\it et al.}}, 2008, \textit{A\&A}, 486, 951
 
\bibitem[S\'anchez-Bl\'azquez {\it et al.} (2006)]{SanchezBlazquez06}
{S\'anchez-Bl\'azquez {\it et al.}} 2006, \textit{MNRAS}, 371, 703

\end{thebibliography}
\end{document}